\documentclass[twocolumn,showpacs,preprintnumbers,amsmath,amssymb,prl,superscriptaddress,floatfix]{revtex4}

\usepackage{graphicx}
\usepackage{dcolumn}
\usepackage{bm}

\providecommand{\KPpnn}{\mbox{${K}^+\to\pi^+\nu\bar\nu$}}
\providecommand{\KPpp}{\mbox{${K}^+\to\pi^+\pi^0$}}
\providecommand{\KPtwo}{\mbox{$K_{\pi2}$}}

\providecommand{\BRSM}{\mbox{$(0.85\pm0.07)\times 10^{-10}$}}  
\providecommand{\BRpnn}{\mbox{$(1.47^{+1.30}_{-0.89})\times 10^{-10}$}} 
\providecommand{\BRthis}{\mbox{$(7.89^{+9.26}_{-5.10})\times10^{-10}$}}
\providecommand{\BRall}{\mbox{$(1.73^{+1.15}_{-1.05})\times 10^{-10}$}}

\providecommand{\TOTbkgd}{\mbox{$0.927\pm0.168{}^{+0.320}_{-0.237}$}}
\providecommand{\TOTbkgdTRIM}{\mbox{$0.93\pm0.17({\rm stat.}){}^{+0.32}_{-0.24}({\rm syst.})$}} 

\begin{document}


\title{New measurement of the \KPpnn\ branching ratio}


\author{A.V.~Artamonov}\affiliation{Institute for High Energy Physics, Protvino, Moscow Region, 142 280, Russia}
\author{B.~Bassalleck}\affiliation{Department of Physics and Astronomy, University of New Mexico, Albuquerque, NM 87131}
\author{B.~Bhuyan}\altaffiliation{Now at Department of Physics, Indian Institute of Technology Guwahati, Guwahati, Assam, 781 039, India.}\affiliation{Brookhaven National Laboratory, Upton, NY 11973}
\author{E.W.~Blackmore}\affiliation{TRIUMF, 4004 Wesbrook Mall, Vancouver, British Columbia, Canada V6T 2A3}
\author{D.A.~Bryman} \affiliation{Department of Physics and Astronomy, University of British Columbia, Vancouver, British Columbia, Canada V6T 1Z1}
\author{S.~Chen} \affiliation{Department of Engineering Physics, Tsinghua University, Beijing 100084, China} \affiliation{TRIUMF, 4004 Wesbrook Mall, Vancouver, British Columbia, Canada V6T 2A3} 
\author{I-H.~Chiang} \affiliation{Brookhaven National Laboratory, Upton, NY 11973}
\author{I.-A.~Christidi}\altaffiliation{Now at Physics Department, Aristotle University of Thessaloniki, Thessaloniki 54124, Greece} \affiliation{Department of Physics and Astronomy, Stony Brook University, Stony Brook, NY 11794}
\author{P.S.~Cooper}\affiliation{Fermi National Accelerator Laboratory, Batavia, IL 60510}
\author{M.V.~Diwan} \affiliation{Brookhaven National Laboratory, Upton, NY 11973}
\author{J.S.~Frank} \affiliation{Brookhaven National Laboratory, Upton, NY 11973}
\author{T.~Fujiwara}\affiliation{Department of Physics, Kyoto University, Sakyo-ku, Kyoto 606-8502, Japan}
\author{J.~Hu} \affiliation{TRIUMF, 4004 Wesbrook Mall, Vancouver, British Columbia, Canada V6T 2A3}
\author{J.~Ives} \affiliation{Department of Physics and Astronomy, University of British Columbia, Vancouver, British Columbia, Canada V6T 1Z1} 
\author{D.E.~Jaffe} \affiliation{Brookhaven National Laboratory, Upton, NY 11973}
\author{S.~Kabe} \affiliation{High Energy Accelerator Research Organization (KEK), Oho, Tsukuba, Ibaraki 305-0801, Japan}
\author{S.H.~Kettell} \affiliation{Brookhaven National Laboratory, Upton, NY 11973}
\author{M.M.~Khabibullin}\affiliation{Institute for Nuclear Research RAS, 60 October Revolution Prospect 7a, 117312 Moscow, Russia}
\author{A.N.~Khotjantsev}\affiliation{Institute for Nuclear Research RAS, 60 October Revolution Prospect 7a, 117312 Moscow, Russia}
\author{P.~Kitching} \affiliation{Centre for Subatomic Research, University of Alberta, Edmonton, Canada T6G 2N5}
\author{M.~Kobayashi} \affiliation{High Energy Accelerator Research Organization (KEK), Oho, Tsukuba, Ibaraki 305-0801, Japan}
\author{T.K.~Komatsubara} \affiliation{High Energy Accelerator Research Organization (KEK), Oho, Tsukuba, Ibaraki 305-0801, Japan}
\author{A.~Konaka} \affiliation{TRIUMF, 4004 Wesbrook Mall, Vancouver, British Columbia, Canada V6T 2A3}
\author{A.P.~Kozhevnikov}\affiliation{Institute for High Energy Physics, Protvino, Moscow Region, 142 280, Russia}
\author{Yu.G.~Kudenko}\affiliation{Institute for Nuclear Research RAS, 60 October Revolution Prospect 7a, 117312 Moscow, Russia} 
\author{A.~Kushnirenko} \altaffiliation{Now at Institute for High Energy Physics, Protvino, Moscow Region, 142 280, Russia.}  \affiliation{Fermi National Accelerator Laboratory, Batavia, IL 60510} 
\author{L.G.~Landsberg}\altaffiliation{Deceased.}\affiliation{Institute for High Energy Physics, Protvino, Moscow Region, 142 280, Russia}
\author{B.~Lewis}\affiliation{Department of Physics and Astronomy, University of New Mexico, Albuquerque, NM 87131}
\author{K.K.~Li}\affiliation{Brookhaven National Laboratory, Upton, NY 11973}
\author{L.S.~Littenberg} \affiliation{Brookhaven National Laboratory, Upton, NY 11973}
\author{J.A.~Macdonald} \altaffiliation{Deceased.} \affiliation{TRIUMF, 4004 Wesbrook Mall, Vancouver, British Columbia, Canada V6T 2A3}
\author{J.~Mildenberger} \affiliation{TRIUMF, 4004 Wesbrook Mall, Vancouver, British Columbia, Canada V6T 2A3}
\author{O.V.~Mineev}\affiliation{Institute for Nuclear Research RAS, 60 October Revolution Prospect 7a, 117312 Moscow, Russia}
\author{M. Miyajima} \affiliation{Department of Applied Physics, Fukui University, 3-9-1 Bunkyo, Fukui, Fukui 910-8507, Japan}
\author{K.~Mizouchi}\affiliation{Department of Physics, Kyoto University, Sakyo-ku, Kyoto 606-8502, Japan}
\author{V.A.~Mukhin}\affiliation{Institute for High Energy Physics, Protvino, Moscow Region, 142 280, Russia}
\author{N.~Muramatsu}\affiliation{Research Center for Nuclear Physics, Osaka University, 10-1 Mihogaoka, Ibaraki, Osaka 567-0047, Japan}
\author{T.~Nakano}\affiliation{Research Center for Nuclear Physics, Osaka University, 10-1 Mihogaoka, Ibaraki, Osaka 567-0047, Japan}
\author{M.~Nomachi}\affiliation{Laboratory of Nuclear Studies, Osaka University, 1-1 Machikaneyama, Toyonaka, Osaka 560-0043, Japan}
\author{T.~Nomura}\affiliation{Department of Physics, Kyoto University, Sakyo-ku, Kyoto 606-8502, Japan}
\author{T.~Numao} \affiliation{TRIUMF, 4004 Wesbrook Mall, Vancouver, British Columbia, Canada V6T 2A3}
\author{V.F.~Obraztsov}\affiliation{Institute for High Energy Physics, Protvino, Moscow Region, 142 280, Russia}

\author{K.~Omata}\affiliation{High Energy Accelerator Research Organization (KEK), Oho, Tsukuba, Ibaraki 305-0801, Japan}
\author{D.I.~Patalakha}\affiliation{Institute for High Energy Physics, Protvino, Moscow Region, 142 280, Russia}
\author{S.V.~Petrenko}\affiliation{Institute for High Energy Physics, Protvino, Moscow Region, 142 280, Russia}
\author{R.~Poutissou} \affiliation{TRIUMF, 4004 Wesbrook Mall, Vancouver, British Columbia, Canada V6T 2A3}
\author{E.J.~Ramberg}\affiliation{Fermi National Accelerator Laboratory, Batavia, IL 60510}
\author{G.~Redlinger} \affiliation{Brookhaven National Laboratory, Upton, NY 11973}
\author{T.~Sato} \affiliation{High Energy Accelerator Research Organization (KEK), Oho, Tsukuba, Ibaraki 305-0801, Japan}
\author{T.~Sekiguchi}\affiliation{High Energy Accelerator Research Organization (KEK), Oho, Tsukuba, Ibaraki 305-0801, Japan}
\author{T.~Shinkawa} \affiliation{Department of Applied Physics, National Defense Academy, Yokosuka, Kanagawa 239-8686, Japan}
\author{R.C.~Strand} \affiliation{Brookhaven National Laboratory, Upton, NY 11973}
\author{S.~Sugimoto} \affiliation{High Energy Accelerator Research Organization (KEK), Oho, Tsukuba, Ibaraki 305-0801, Japan}
\author{Y.~Tamagawa} \affiliation{Department of Applied Physics, Fukui University, 3-9-1 Bunkyo, Fukui, Fukui 910-8507, Japan}
\author{R.~Tschirhart}\affiliation{Fermi National Accelerator Laboratory, Batavia, IL 60510}
\author{T.~Tsunemi}\altaffiliation{Now at Department of Physics, Kyoto University, Sakyo-ku, Kyoto 606-8502, Japan.}\affiliation{High Energy Accelerator Research Organization (KEK), Oho, Tsukuba, Ibaraki 305-0801, Japan}
\author{D.V.~Vavilov}\affiliation{Institute for High Energy Physics, Protvino, Moscow Region, 142 280, Russia}
\author{B.~Viren}\affiliation{Brookhaven National Laboratory, Upton, NY 11973}
\author{Zhe~Wang} \affiliation{Department of Engineering Physics, Tsinghua University, Beijing 100084, China} \affiliation{Brookhaven National Laboratory, Upton, NY 11973} 
\author{N.V.~Yershov}\affiliation{Institute for Nuclear Research RAS, 60 October Revolution Prospect 7a, 117312 Moscow, Russia}
\author{Y.~Yoshimura} \affiliation{High Energy Accelerator Research Organization (KEK), Oho, Tsukuba, Ibaraki 305-0801, Japan}
\author{T.~Yoshioka}\affiliation{High Energy Accelerator Research Organization (KEK), Oho, Tsukuba, Ibaraki 305-0801, Japan}
\collaboration{E949 Collaboration}\noaffiliation

\date{\today} 

\preprint{BNL-81421-2008-JA}
\preprint{FERMILAB-PUB-08-301-CD-E}
\preprint{KEK/2008-25}
\preprint{TRIUMF/TRI-PP-08-04}
\preprint{TUHEP-EX-08-03}





\begin{abstract}
 Three 
events for the decay \KPpnn\ have been observed in the pion 
momentum region below the \KPpp\ peak, 
$140 < P_\pi < 199\ {\rm MeV}/c$, with an estimated background 
of \TOTbkgdTRIM\ events. Combining 
this observation with previously reported results yields
a branching ratio of ${\cal B}(\KPpnn) = \BRall$ 
consistent with the standard model prediction.

\end{abstract}

\pacs{13.20.-v, 12.15.Hh} 

\maketitle

 The decay \KPpnn\  is among a handful of hadronic processes for which the decay rate can be 
accurately  predicted in the standard model  (SM) owing to knowledge of the transition
matrix element from similar processes and minimal long-distance effects~\cite{Buras:2004uu,Mescia:2007kn}. 
The small  predicted branching ratio,
 ${\cal B}(\KPpnn)$ = \BRSM\ \cite{ref:TH2},  and the fact that
this decay is a flavor-changing neutral current process makes it a 
sensitive probe of a wide range of 
new physics effects~\cite{Buras:2004uu}. Previous studies of
this decay by experiment E787 at Brookhaven National Laboratory and its upgraded extension
E949 have measured ${\cal B}(\KPpnn)=\BRpnn$ based on the observation of three events 
in a sample of $7.7\times10^{12}$ 
${K}^+$ decays at rest 
with a total expected background of $0.44\pm0.05$ events in the pion 
momentum region $211<P_\pi<229\ {\rm MeV}/c$  above the 
${K}^+\to\pi^+\pi^0$ ({\KPtwo}) peak (pnn1)~\cite{ref:pnn1_PRL,ref:pnn1_PRD}.
 E787 set a consistent limit
of $<22\times 10^{-10}$ at 90\% C.L. based on one candidate 
in a sample of $1.7\times10^{12}$ stopped ${K}^+$ decays 
with an expected 
background of $1.22\pm0.24$ events   in the momentum
region $140 < P_\pi < 195\ {\rm MeV}/c$  below the \KPtwo\ 
peak (pnn2)~\cite{ref:pnn2_PLB,ref:pnn2_PRD}.

 In this Letter we report the results of a search for \KPpnn\ below the \KPtwo\ peak (pnn2) 
using $1.7\times 10^{12}$ stopped ${K}^+$ decays obtained with E949 as well as
the final results on ${\cal B}(\KPpnn)$ from  E949 data combined with E787
data.

Identification of  \KPpnn\ decays 
relies on detection of an incoming kaon, its decay at rest 
and an outgoing pion with no coincident 
detector activity. 
The E949 apparatus and analysis of the data in the pnn1 region have been
described elsewhere~\cite{ref:pnn1_PRD}. 
In this letter, 
we emphasize 
the apparatus
and analysis features most relevant for pnn2. 

Incoming kaons were identified by a {\v C}erenkov counter and two proportional wire
chambers before being slowed by an 11.1 cm thick BeO degrader  and an 
 active degrader, 
passing through a beam hodoscope and 
stopping in the scintillating fiber target. 
Typically $1.6\times 10^6\ {K}^+/{\rm s}$ entered the  target 
during a 2.2 s spill with a ${K}^+/\pi^+$ ratio of 3. 
The active degrader had 
39 copper disks (2.2 mm thick) interleaved with  40 layers of 2 mm plastic scintillator
divided into 12 azimuthal segments. Scintillation light
from each segment was transported via wavelength shifting fibers to a photomultiplier
tube (PMT) that was read  out by time-to-digital convertors (TDCs), 
analog-to-digital convertors (ADCs) and 
GaAs CCD waveform digitizers  (CCDs) sampling at 500 MHz~\cite{ref:CCD}.
 The active degrader was capable of providing measurements
of the incoming beam particle and activity concident with ${K}^+$ decay 
in the target.
 The target consisted of 413 scintillating
fibers  (5 mm square and 3.1 m long) packed into a 12 cm diameter cylinder. 
Each 5 mm fiber was connected to a PMT and read out by
TDCs, ADCs and CCDs in order to record 
 activity in the target coincident with both  the incoming kaon 
and  the outgoing pion. 

 The momentum and trajectory 
of the outgoing  $\pi^+$
were measured in a drift chamber~\cite{ref:UTC}. The outgoing pion 
came to rest 
in a range stack 
of 19 layers of
plastic scintillator with 24 segments in azimuth. PMTs on each end of the
scintillator were read out by TDCs, ADCs and 500-MHz transient
digitizers~\cite{ref:TD} and enabled measurement of the pion range ($R_\pi$) and 
kinetic energy ($E_\pi$) as well as the $\pi^+\to\mu^+\to e^+$ decay sequence.

 The barrel veto  
 calorimeters of 16.6 radiation lengths (r.l.) at
normal incidence provided photon detection over 2/3 
of $4\pi$ sr solid angle. 
Photon detection over the remaining $1/3$ of $4\pi$ sr solid angle 
was provided by a variety of calorimeters in the region from
$10^\circ$ to $45^\circ$ of the beam axis
with a total thickness from $7$ to $15$ r.l.~\cite{ref:pnn1_PRD,Chiang:1995ar,Komatsubara:1997rq,Mineev:2002cu}. 
More extensive use was made by this analysis than the pnn1 analysis 
of the photon detection capabilities of the active degrader (6.1 r.l.) and the target (7.3 r.l.) 
that occupied the region within 
 $10^\circ$ of the beam axis.

This pnn2 analysis was able to increase the signal acceptance by 40\% and
maintain the same background rate per stopped ${K}^+$ 
as the previous analysis~\cite{ref:pnn2_PRD} 
thanks to improved background rejection 
primarily due to 
the addition of the active degrader and augmentation of the barrel veto by 2.3 r.l. for E949. 
In addition, the improved knowledge of the background contributions allowed
the signal region to be divided into nine sub-regions (``cells''),  with relative
signal-to-background levels differing by a factor of 4,  that
were used in the likelihood method~\cite{ref:Junk} to 
determine ${\cal B}(\KPpnn)$.

 To avoid a possible bias, 
we employed a ``blind analysis'' technique~\cite{ref:pnn1_PRD} in which the
signal region was not examined until all 
selection criteria (``cuts'') for signal had been established, 
the estimates of all 
backgrounds  completed and acceptance of all cells determined. 
Two uncorrelated cuts with significant rejection were
developed for most backgrounds. 
After imposing 
basic event quality cuts, 
inversion of one of the pair 
of cuts could then be used to select a background-enriched
data sample containing $N$ events. Inversion of the complementary 
cut selected a data sample on which the rejection ${\cal R}$ of the
first cut could be measured. The background was estimated 
as $N/({\cal R}-1)$. We ensured unbiased background estimates 
by dividing the data into one-third and two-thirds samples
chosen uniformly from the entire data set. Selection criteria
were determined with the one-third sample and background levels were 
measured  from the two-thirds sample. In contrast to the analysis of the 
pnn1 region, some backgrounds did 
not have sufficiently distinct 
characteristics to permit isolation by cut inversion of a
pure background sample and permit a measurement of ${\cal R}$ 
with the data. For these backgrounds, ${\cal R}$ 
was estimated with simulated data as described below.

Table~\ref{tab:1} summarizes the estimated background levels.
The largest background was due to \KPtwo\ decays in which the
$\pi^+$ scatters in the target, losing energy and 
obscuring the directional correlation with the photons from the 
$\pi^0$ decay that would otherwise be detected in the barrel veto. 
Two cuts that suppressed this background were 1) identification 
of $\pi^+$ scattering and 2) detection of the photons from $\pi^0$ decay. 
Pion scattering
was identified 
by kinks in the pattern of target fibers attributed to the pion, 
by tracks that did not point back to the fiber containing the ${K}^+$ decay, 
by energy deposits inconsistent with an outgoing pion or 
by unexpected energy deposits  at the time of the  pion in fibers traversed by the kaon. 
The target pulse-shape 
cut identified the 
latter signature by performing a least-squares fit to the 
CCD samples to identify the pulses due to activity coincident with the kaon or 
pion~\cite{ref:pnn2_PRD}. 
The uncertainty in the \KPtwo\ target-scatter background had 
comparable statistical and systematic contributions. 
The systematic uncertainty was determined by the range of photon veto 
rejection
values measured on samples of \KPtwo\ scatter events selected by different
scattering signatures in the target or in different $\pi^+$ kinematic 
regions~\cite{ref:footnote}. 
There was also a much smaller background from \KPtwo\ 
due to scattering in the range stack that was similarly identified by 
the energy deposits and pattern of range stack counters
attributed to the  track.

\begin{table}
\begin{center}
\begin{tabular}{l|l}
\hline
Process           & Background events       \\ \hline
\KPtwo\ target-scatter      &$0.619\pm0.150{}^{+0.067}_{-0.100}$      \\
\KPtwo\ range-stack-scatter      &$0.030\pm0.005\pm0.004$                \\
$K_{\pi2\gamma}$   &$0.076\pm0.007\pm0.006$               \\
$K_{e4}$          &$0.176\pm0.072{}^{+0.233}_{-0.124}$     \\
Charge-exchange                &$0.013\pm{0.013}{}^{+0.010}_{-0.003}$ \\
Muon               &$0.011\pm0.011$                     \\
Beam               &$0.001\pm0.001$                     \\ \hline
Total              &\TOTbkgd\                           \\ \hline
\end{tabular}
\caption{\label{tab:1}
Summary of the estimated number of events in the signal region 
from each background component. 
Each component is described in the text.}
\end{center}
\end{table}


 Additional backgrounds included
${K}^+\to\pi^+\pi^- e^+\nu$ ($K_{e4}$), 
 ${K}^+\to\pi^+\pi^0\gamma$ ($K_{\pi2\gamma}$), 
${K}^+\to\mu^+\nu$, ${K}^+\to\mu^+\nu\gamma$ and ${K}^+\to\pi^0\mu^+\nu$ (muon), 
scattered beam pions (beam) and 
$\pi^+$ resulting from ${K}^+$ charge-exchange
reactions dominated by 
${K}^0_{\rm L}\to\pi^+\mu^-\bar\nu$. 
Simulated data were used to estimate the
rejection ${\cal R}$ of the cuts that suppress $K_{e4}$, $K_{\pi2\gamma}$ and charge-exchange
backgrounds. The $K_{e4}$ and $K_{\pi2\gamma}$ backgrounds 
could not be distinguished from the larger 
{\KPtwo}-scatter background based solely on the $\pi^+$ track, and it was not
possible to isolate a sufficiently pure, statistically significant sample of 
charge-exchange events on which to measure ${\cal R}$.

The $K_{e4}$ 
process forms a background when the $\pi^-$ and $e^+$ interact in the 
target without leaving a detectable trace.  Positron interactions 
were well-modelled in our EGS4-based simulation~\cite{ref:EGS} and 
we used the $\pi^-$ energy deposition spectrum in scintillator measured
previously in E787~\cite{ref:piminus} to model $\pi^-$ absorption. 
We assessed the systematic uncertainty in the $K_{e4}$ background by 
varying the threshold of cuts on the energy deposited 
in the target fibers at the time of the pion.
The kinematics cuts defining the 
signal region were
$140<P_\pi<199\ {\rm MeV}/c$, 
$60<E_\pi<100.5\ {\rm MeV}$ and $12<R_\pi<28\ {\rm cm}$. 
We defined a sub-region 
$165<P_\pi<197\ {\rm MeV}/c$, 
$72<E_\pi<100\ {\rm MeV}$ and $17<R_\pi<28\ {\rm cm}$ where 
the lower  and upper limits were chosen to suppress the $K_{e4}$ background that peaks 
near $160\ {\rm MeV}/c$ and the tail of the \KPtwo\ peak, respectively.

The rejection of the $K_{\pi2\gamma}$ background 
was calculated using a combination of simulated \KPtwo\ and $K_{\pi2\gamma}$ 
events and \KPtwo\ data events. The additional photon veto rejection due to the
radiative photon was calculated from the photon distribution 
in simulated events and the rejection power of single photons as a
function of angle and energy evaluated with \KPtwo\  data~\cite{ref:Kentaro}. 

Measurements of $K_{\rm S}^0\to\pi^+\pi^-$ decay from 
the ${K}^+$ charge-exchange reaction were 
used as
input to  simulate charge-exchange events~\cite{ref:pnn1_PRD}. The requirement on
the delayed coincidence between the reconstructed kaon 
and pion candidates provided suppression of charge-exchange background as the
emitted $\pi^+$ was required to originate within the fiducial region of the
target. The systematic uncertainty was assessed with the same methodology 
as the $K_{e4}$ background. 

The  muon and beam backgrounds were estimated entirely from
data and were very small.
As previous analyses had shown 
the muon background to be small~\cite{ref:pnn2_PLB,ref:pnn2_PRD}, 
the transient-digitizer-based cuts on $\pi^+\to\mu^+\to e^+$ 
identification were loosened to gain about $10\%$ in acceptance. 
The total acceptance of the signal region  was 
$(1.37\pm0.14)\times10^{-3}$. 

In order to explore and verify the reliability of the background
estimates, we examined three distinct data regions just outside
the signal region by loosening the photon veto ($PV_n$) or target pulse-shape ($CCD_n$) cut.
Each of the two regions, 
$PV_1$ and $CCD_1$, were immediately adjacent to the signal
region while a third region $PV_2$, adjacent to $PV_1$, 
was defined by further loosening of the photon veto cut.
The number of expected and observed events and the probability of the observation 
are given in Table~\ref{tab:2}. 
The 5\% 
probability for the
regions nearest the signal region may have indicated that the background was over-estimated. 
Given the inability to cleanly isolate each background component by cut inversion, 
some contamination (i.e. events due to backgrounds from other sources) 
is possible and would generally inflate the background estimates. 
Re-evaluation of  the probabilities at the lower limit of the systematic 
uncertainties~\cite{ref:footnote} 
gave 14\% 
for the two closest regions and demonstrated that the
assigned systematic uncertainties were reasonable.

\begin{table}
\begin{center}
\begin{tabular}{c|c|c|c|c}
\hline
  Region       & $N_{\rm E}$           & $N_{\rm O}$ & ${\cal P}(N_{\rm O}; N_{\rm E})$ & Combined \\ \hline
$CCD_1$        &$0.79^{+0.46}_{-0.51}$    &  0           & 0.45 [0.29,0.62]                         & NA       \\
$PV_1$         &$9.09^{+1.53}_{-1.32}$    &  3           & 0.02 [0.01,0.05]                         & 0.05 [0.02,0.14] \\
$PV_2$         &$32.4^{+12.3}_{-8.1}$     & 34           & 0.61 [0.05,0.98]                         & 0.14 [0.01,0.40] \\
\hline
\end{tabular}
\caption{\label{tab:2}Comparison of the expected $(N_{\rm E})$ and observed $(N_{\rm O})$ number of background 
events in three regions $CCD_1$, $PV_1$ and $PV_2$ outside 
the signal region.
The central value of $N_{\rm E}$ is given along
with the combined statistical and systematic uncertainties. 
${\cal P}(N_{\rm O}; N_{\rm E})$ is
the probability of observing $N_{\rm O}$ events or fewer when $N_{\rm E}$ events are expected. The 
rightmost column ``Combined'' gives the probability of the combined observation in that region and the region(s) 
of the preceding row(s). The numbers in square brackets are the probabilities re-evaluated 
at the upper and lower bounds of the uncertainty on $N_{\rm E}$~\protect\cite{ref:footnote}. 
}
\end{center}
\end{table}

\begin{figure}
\begin{center}
\includegraphics[width=1.1\linewidth]{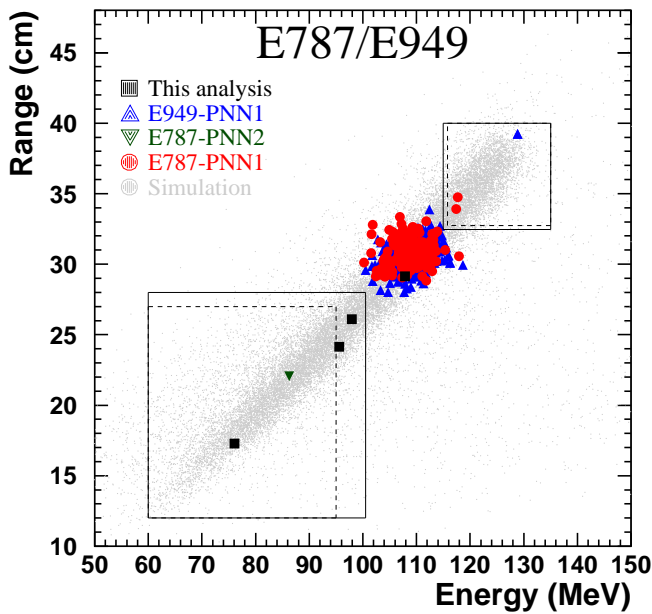} 
\caption{\label{fig:1} 
Kinetic energy {\sl vs.} 
range of all 
events passing all other 
cuts.
The squares represent the events selected by this analysis.
The circles and upward-pointing triangles represent the events 
selected by the E787 and E949 pnn1 analyses, respectively.
The downward-pointing triangles represent the 
events selected by the E787 pnn2 analyses. 
 The solid (dashed) lines 
represent the limits of the pnn1 and pnn2 signal regions for 
the E949 (E787) analyses. Despite the smaller signal region in $E_\pi$ {\sl vs.} $R_\pi$, 
the pnn1 analyses were 4.2 times more sensitive than the pnn2 analyses. 
The points near $E_\pi = 108$ MeV were \KPtwo\ decays that survived the photon veto cuts
and were predominantly from the pnn1 analyses due to the higher sensitivity and
the less stringent photon veto cuts.  
The light gray points are simulated \KPpnn\ events that would be
accepted by our trigger.
}
\end{center}
\end{figure}

After completion of the background studies, the signal region was examined
and three candidates were found.
The energy {\sl vs.} range for these observed candidates 
is shown in Figure~\ref{fig:1} along with the results of previous 
E787~\cite{ref:pnn2_PLB,ref:pnn2_PRD} and 
E949~\cite{ref:pnn1_PRL,ref:pnn1_PRD}  analyses.
 From these three new events 
alone, ${\cal B}(\KPpnn) = \BRthis$
was calculated using the likelihood method~\cite{ref:Junk} 
assuming the SM spectrum and taking into account the
uncertainties in the background and acceptance measurements~\cite{ref:Joss}. 
When combined with 
the results of previous E787 and E949   analyses, 
we 
found ${\cal B}(\KPpnn) = {\BRall}$. 
The signal-to-background (S/B) ratios
for the three events 
are 0.20, 0.42 and 0.47~\cite{ref:three}, which can be compared with 
the S/B = 0.20 for the previous pnn2 candidate~\cite{ref:pnn2_PLB} and 
with the S/B = 59, 8.2 and 1.1 for the pnn1 events~\cite{ref:pnn1_PRL} 
assuming ${\cal B}(\KPpnn) = 1.73\times10^{-10}$.
In this 
analysis, a candidate in 
the best (worst) cell would have had S/B=0.84 (0.20).
The probability that the three new events
were due to background only, 
given the estimated background in each cell, is 0.037. 
The probability that 
all seven \KPpnn\ events 
were due to background is 0.001. In summary, these observations 
imply a \KPpnn\ branching ratio consistent with SM 
 expectations. 

We 
acknowledge the dedicated effort of the technical staff supporting
E949, the Brookhaven C-A Department,
and  the  contributions made by colleagues who
participated in  E787.  This research
was supported in part by the U.S. Department of Energy, 
the Ministry of Education, Culture, Sports, Science and Technology of
Japan through the Japan-U.S. Cooperative Research Program in High
Energy Physics and under Grant-in-Aids for Scientific Research, 
the Natural Sciences and Engineering Research Council and the National Research Council of Canada, 
the program for New Century Excellent Talents in University from the Chinese Ministry of Education, 
the Russian Federation State Scientific Center Institute for High Energy Physics, 
and 
the Ministry of Industry, Science and
New Technologies of the Russian Federation.

\bibliography{aps_pnn2_v7}

\begin{thebibliography}{20}
\expandafter\ifx\csname natexlab\endcsname\relax\def\natexlab#1{#1}\fi
\expandafter\ifx\csname bibnamefont\endcsname\relax
  \def\bibnamefont#1{#1}\fi
\expandafter\ifx\csname bibfnamefont\endcsname\relax
  \def\bibfnamefont#1{#1}\fi
\expandafter\ifx\csname citenamefont\endcsname\relax
  \def\citenamefont#1{#1}\fi
\expandafter\ifx\csname url\endcsname\relax
  \def\url#1{\texttt{#1}}\fi
\expandafter\ifx\csname urlprefix\endcsname\relax\def\urlprefix{URL }\fi
\providecommand{\bibinfo}[2]{#2}
\providecommand{\eprint}[2][]{\url{#2}}

\bibitem[{\citenamefont{Buras et~al.}(2008)\citenamefont{Buras, Schwab, and
  Uhlig}}]{Buras:2004uu}
\bibinfo{author}{\bibfnamefont{A.~J.} \bibnamefont{Buras}},
  \bibinfo{author}{\bibfnamefont{F.}~\bibnamefont{Schwab}}, \bibnamefont{and}
  \bibinfo{author}{\bibfnamefont{S.}~\bibnamefont{Uhlig}},
  \bibinfo{journal}{Rev. Mod. Phys.} \textbf{\bibinfo{volume}{80}},
  \bibinfo{pages}{965} (\bibinfo{year}{2008}), \eprint{hep-ph/0405132}.

\bibitem[{\citenamefont{Mescia and Smith}(2007)}]{Mescia:2007kn}
\bibinfo{author}{\bibfnamefont{F.}~\bibnamefont{Mescia}} \bibnamefont{and}
  \bibinfo{author}{\bibfnamefont{C.}~\bibnamefont{Smith}},
  \bibinfo{journal}{Phys. Rev.} \textbf{\bibinfo{volume}{D76}},
  \bibinfo{pages}{034017} (\bibinfo{year}{2007}), \eprint{arXiv:0705.2025}.

\bibitem[{\citenamefont{Brod and Gorbahn}(2008)}]{ref:TH2}
\bibinfo{author}{\bibfnamefont{J.}~\bibnamefont{Brod}} \bibnamefont{and}
  \bibinfo{author}{\bibfnamefont{M.}~\bibnamefont{Gorbahn}},
  \bibinfo{journal}{Phys. Rev.} \textbf{\bibinfo{volume}{D78}},
  \bibinfo{pages}{034006} (\bibinfo{year}{2008}),
  \bibinfo{note}{{arXiv:0805.4119. The uncertainty in the prediction is
  dominated by the uncertainty in the elements of the CKM matrix.}}

\bibitem[{\citenamefont{Anisimovsky et~al.}(2004)}]{ref:pnn1_PRL}
\bibinfo{author}{\bibfnamefont{V.~V.} \bibnamefont{Anisimovsky}}
  \bibnamefont{et~al.}, \bibinfo{journal}{Phys. Rev. Lett.}
  \textbf{\bibinfo{volume}{93}}, \bibinfo{pages}{031801}
  (\bibinfo{year}{2004}), \eprint{hep-ex/0403036}.

\bibitem[{\citenamefont{Adler et~al.}(2008)}]{ref:pnn1_PRD}
\bibinfo{author}{\bibfnamefont{S.}~\bibnamefont{Adler}} \bibnamefont{et~al.},
  \bibinfo{journal}{Phys. Rev.} \textbf{\bibinfo{volume}{D77}},
  \bibinfo{pages}{052003} (\bibinfo{year}{2008}), \eprint{arXiv:0709.1000}.

\bibitem[{\citenamefont{Adler et~al.}(2002)}]{ref:pnn2_PLB}
\bibinfo{author}{\bibfnamefont{S.}~\bibnamefont{Adler}} \bibnamefont{et~al.},
  \bibinfo{journal}{Phys. Lett.} \textbf{\bibinfo{volume}{B537}},
  \bibinfo{pages}{211} (\bibinfo{year}{2002}), \eprint{hep-ex/0201037}.

\bibitem[{\citenamefont{Adler et~al.}(2004)}]{ref:pnn2_PRD}
\bibinfo{author}{\bibfnamefont{S.}~\bibnamefont{Adler}} \bibnamefont{et~al.},
  \bibinfo{journal}{Phys. Rev.} \textbf{\bibinfo{volume}{D70}},
  \bibinfo{pages}{037102} (\bibinfo{year}{2004}), \eprint{hep-ex/0403034}.

\bibitem[{\citenamefont{Bryman et~al.}(1997)}]{ref:CCD}
\bibinfo{author}{\bibfnamefont{D.~A.} \bibnamefont{Bryman}}
  \bibnamefont{et~al.}, \bibinfo{journal}{Nucl. Instrum. Meth.}
  \textbf{\bibinfo{volume}{A396}}, \bibinfo{pages}{394} (\bibinfo{year}{1997}),
  \bibinfo{note}{{a CCD is a charge-coupled device.}}

\bibitem[{\citenamefont{Blackmore et~al.}(1998)}]{ref:UTC}
\bibinfo{author}{\bibfnamefont{E.~W.} \bibnamefont{Blackmore}}
  \bibnamefont{et~al.}, \bibinfo{journal}{Nucl. Instrum. Meth.}
  \textbf{\bibinfo{volume}{A404}}, \bibinfo{pages}{295} (\bibinfo{year}{1998}).

\bibitem[{\citenamefont{Atiya et~al.}(1989)\citenamefont{Atiya, Ito, Haggerty,
  Ng, and Sippach}}]{ref:TD}
\bibinfo{author}{\bibfnamefont{M.}~\bibnamefont{Atiya}},
  \bibinfo{author}{\bibfnamefont{M.}~\bibnamefont{Ito}},
  \bibinfo{author}{\bibfnamefont{J.}~\bibnamefont{Haggerty}},
  \bibinfo{author}{\bibfnamefont{C.}~\bibnamefont{Ng}}, \bibnamefont{and}
  \bibinfo{author}{\bibfnamefont{F.~W.} \bibnamefont{Sippach}},
  \bibinfo{journal}{Nucl. Instrum. Meth.} \textbf{\bibinfo{volume}{A279}},
  \bibinfo{pages}{180} (\bibinfo{year}{1989}).

\bibitem[{\citenamefont{Chiang et~al.}(1995)}]{Chiang:1995ar}
\bibinfo{author}{\bibfnamefont{I.~H.} \bibnamefont{Chiang}}
  \bibnamefont{et~al.}, \bibinfo{journal}{IEEE Trans. Nucl. Sci.}
  \textbf{\bibinfo{volume}{42}}, \bibinfo{pages}{394} (\bibinfo{year}{1995}).

\bibitem[{\citenamefont{Komatsubara et~al.}(1998)}]{Komatsubara:1997rq}
\bibinfo{author}{\bibfnamefont{T.~K.} \bibnamefont{Komatsubara}}
  \bibnamefont{et~al.}, \bibinfo{journal}{Nucl. Instrum. Meth.}
  \textbf{\bibinfo{volume}{A404}}, \bibinfo{pages}{315} (\bibinfo{year}{1998}).

\bibitem[{\citenamefont{Mineev et~al.}(2002)}]{Mineev:2002cu}
\bibinfo{author}{\bibfnamefont{O.}~\bibnamefont{Mineev}} \bibnamefont{et~al.},
  \bibinfo{journal}{Nucl. Instrum. Meth.} \textbf{\bibinfo{volume}{A494}},
  \bibinfo{pages}{362} (\bibinfo{year}{2002}), \eprint{physics/0207033}.

\bibitem[{\citenamefont{Junk}(1999)}]{ref:Junk}
\bibinfo{author}{\bibfnamefont{T.}~\bibnamefont{Junk}}, \bibinfo{journal}{Nucl.
  Instrum. Meth.} \textbf{\bibinfo{volume}{A434}}, \bibinfo{pages}{435}
  (\bibinfo{year}{1999}), \eprint{hep-ex/9902006}.

\bibitem[{ref({\natexlab{a}})}]{ref:footnote}
\bibinfo{note}{This method of assigning systematic uncertainty was intended to
  define a range that included the actual value of the background.}

\bibitem[{\citenamefont{Nelson et~al.}(1985)\citenamefont{Nelson, Hirayama, and
  Rogers}}]{ref:EGS}
\bibinfo{author}{\bibfnamefont{W.~R.} \bibnamefont{Nelson}},
  \bibinfo{author}{\bibfnamefont{H.}~\bibnamefont{Hirayama}}, \bibnamefont{and}
  \bibinfo{author}{\bibfnamefont{D.~W.~O.} \bibnamefont{Rogers}}
  (\bibinfo{year}{1985}), \bibinfo{note}{{SLAC}-0265}.

\bibitem[{\citenamefont{Ardebili}(1995)}]{ref:piminus}
\bibinfo{author}{\bibfnamefont{M.}~\bibnamefont{Ardebili}}, Ph.D. thesis,
  \bibinfo{school}{{Princeton University}} (\bibinfo{year}{1995}),
  \bibinfo{note}{{UMI-95-27860}}.

\bibitem[{\citenamefont{Mizouchi}(2006)}]{ref:Kentaro}
\bibinfo{author}{\bibfnamefont{K.}~\bibnamefont{Mizouchi}}, Ph.D. thesis,
  \bibinfo{school}{{Kyoto University}} (\bibinfo{year}{2006}).

\bibitem[{\citenamefont{Ives}()}]{ref:Joss}
\bibinfo{author}{\bibfnamefont{J.}~\bibnamefont{Ives}}, \bibinfo{note}{{Ph.D.
  thesis, University of British Columbia, to be published.}}

\bibitem[{ref({\natexlab{b}})}]{ref:three}
\bibinfo{note}{The kinetic energies of the events with S/B of 0.20, 0.42 and
  0.47 were 76.1, 97.9 and 95.6 MeV, respectively.}

\end{thebibliography}
\end{document}